# The complex social network from The Lord of The Rings

Mauricio Aparecido Ribeiro, Roberto Antonio Vosgerau, Maria Larissa Pereira Andruchiw, Sandro Ely de Souza Pinto

Physics department, Universidade Estadual de Ponta Grossa, Ponta Grossa, Paraná, Brazil

## Abstract

Studies of social structures has been grown on the last years, because its sharing form and content creation attracted the public in general. Such structures are observed, as an example, in literary pieces. A featured author is J.R.R. Tolkien, with his books that describe a fictional world and its inhabitants. These books bring a narrative of the creation of the Middle-Earth and all of its mythology. His main pieces are: "The Silmarillion", "The Hobbit" and "The Lord of the rings", The objective of this article is the analysis of the social structures emerging of the conjunction of these works, where the social relations are described by the reference criteria, shared events and direct bonds, with the major centrality measures together with the structural entropy of first order. Enabling the doing of an analogy with the canonic ensemble of the mechanics statistics and enabling analyzing the degree of homogeneity of the bonds between the formed communities.

## Introduction

The theory of complex networks is widely used for modeling structures found in nature, named real networks [1]. An example of real networks are social structures, which are formed by the interaction of individuals [2]. We can identify as an example of social structure, a classroom, where students interact in several ways. The interaction among individuals it's named social relationship. Another example is discussion communities in the internet, like Facebook, Twitter and others that create this social bond of interaction in virtual form, increasing the possibility of interaction among people. [3], [4].

Nevertheless, we can observe that standards in the social type emerge from literary works, more specifically mythological works. The works with mythological character introduce interactions between the real world and the world of fictional beings, and it was shown that these networks are the complex network type [5], [6], [7], [8], [9]. For the formation of these structures we need to define what types of relationships are formed in order that have social interaction among individuals [10]. So, we define the following criterion for social relations:

We define the following criteria for social relations as:

- Allusion to the social relationship: When in dialogue the characters describe another person with minimal details is response to the social relations involved.

- Sharing events: When characters are in the same location and in the same time, showing the action participation.

- Direct Interaction: characters that interact directly

Here, we consider the declaration of a social relationship among the chacacters, as the pioneering work of Coleman [11].                                              Some works of fiction based on various mythologies and heroics acts belongs to the writer J. R. R. Tolkien [12]. These works report the creation of middle earth and the events and realizations of its inhabitant. This work of Tolkien is based on Greek mythology and scandinavian, thus, creating a unic world.Tthe world created by Tolkien, denominated Arda, it has a beginning that is reported in the work "The Silmarillion", where God Ilúvatar with his symphony of creation with its Ainur, architects everything. Other works as ""The Hobbit" and "The Lord Of The Rings" report the realizations and epic battles with its respective heroes [13]. In "The Hobbit", describes the misadventures of the hobbit Bilbo Bolseiro Who travels to Erebor with the dwarves to conquer it as "Smaug", and with these misadventures, starting the events of The Lord Of The Rings [14]. The work "The Lord Of The Rings" it is divided into three parts, the first one called The Fellowship of the Ring, that report the company formed by Elrond in Valfenda to support Frodo in the power rind destruction. The second work, The Two Towers, reports the disruption of the Fellowship of the Ring and also the events that occur in the towers of Orthanc, in Isengard, and Barad- Dúr, in Mordor. The third work, The Return of the King, describe the battles of Gondor and Rohan, and also the finalization of the work with the destruction of the power ring [15].

These works of J.R.R. Tolkien not only influenced literature, but also RPG games, film and music.  Each of the Tolkien's characters has a genealogic three and carries with him all the culture of its people, as well as its characteristic language, relatives and social relations of the most diverse kinds along the plot.

The main purpose of this article it to deal with a social network from interdisciplinary perspective combining literature, mathematics and physics, also the degree of centrality measures and its distribution,  betweenness centrality, assortativity, transitivity and the average  length path which are classic measures in the characterization of networks. We also analyse the formation of communities with the WalkTrap Comunities algorithm, that use the random walk. We also apply the theory of the canonical ensemble, where, can be calculate the structural entropy. Structural entropy allows measuring the degree of complexity generated by the links that the vertices exert with its adjacent.

## Results and Discussions

Figure 1 represents the social network  with  allusion relations to social relationship, event sharing and direct dial among the characters from Tolkien' works. This structure presents 618 characters ( vertex) and 19462 social relationship ( edges).  The degree distribution of the kind power law with and exponential cut, given by equation 2 that allows the correct adjustment for vertices with high probability of connection, where and anomaly occurs, being characterized as network hubs. In the literary structure, are the vertices that represent the deities and mythological beings, because over the reports a large number of characters know them.  Figure 2 illustrates the degree of distribution to the respctive adjustment. The quality of the adjustment was analyzed by Person's coefficient. ( $R^2$= 0.900672). With α= $10^{-5}$, β =−508.64, γ = 0.07068 e ν = 0.06563.

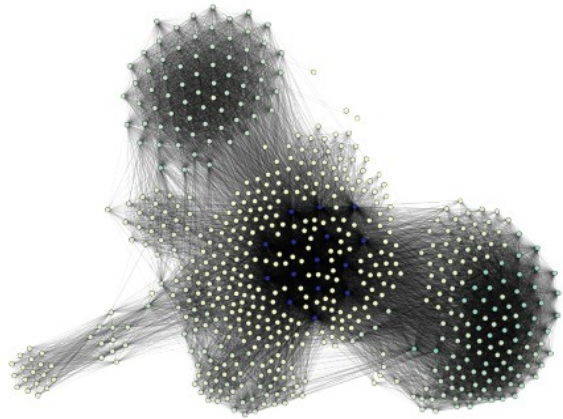

**Figure 1:** Social network of the books "The Silmarillion, The Hobbit and The Lord Of The Rings unique volume". With N= 618 (characters), L= 19462 (social relationship). The blue vertices are the one Who has the highest connection to the network.

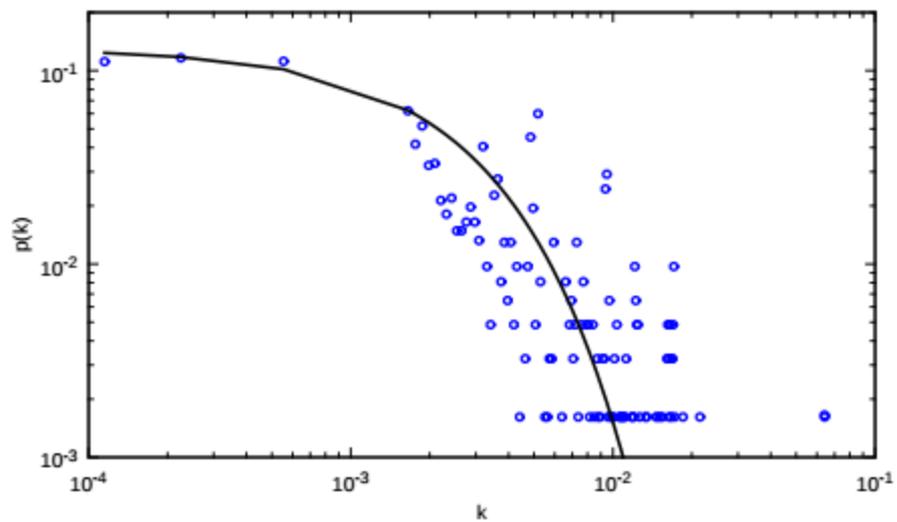

**Figure 2:** Degree distribution of social structure with their respective adjustment given by the equation 2, the quality of the adjustment it is given by the coefficient of Pearson $R^2$

In table 1 are presented the structural measures of the network formed with the deities and without the deities

**Table 1:** Values of the structural properties of social network formed by characters from the books (The Silmarillion, The Hobbit, The Lord Of The Rings unique volume) from J.R.R Tolkien. Where <k> is the average degree, (T) transitivity, (A) assortativity, (< B > ) average betweenness centrality, (D) medium way distance, (N) number of vertices and (E) number of edges.

| Properties | Complete social structure | Social structure without the deities |
|---|---|---|
| <k> | 63 | 48 |
| T | 0,318 | 0,377 |
| A | -0,367 | -0,378 |
| (<B>) | 304,894 | 307,816 |
| (D) | 1,988 | 3,961 |
| (N) | 618 | 608 |
| (E) | 19462 | 14798 |

The network presented a disassortative character, in other words, characters with highest connection tends to connect with the ones that has lowest connection. This happens due to the works had a description of the "gods" (valar) from Middle- Earth. The interaction of the gods to the characters happens in a way that the distance of medium way be low. This evidences the omnipresent character of divinity, as shown in [7]. In the cited reference, when the mythological characters of the network are removed, the quantity D rises from ≈ 2 for ≈4.0.

The measure of transitivity, which is referring to triangles formed in the structure, indicates the relation of sharing the events and allusion, because the transitivity calculated indicates that the characters tend to form cohesive groups. The tendency of formation of these cohesive groups allow the calculation of the communities on the network. This separation of the communities occurred with the optimization of modularity function. The optimization occurred with Walktrap Communities algorithm that uses the random walk, identifying 5 communities as shown in figure 4. Once identified the communities of the social structure, it is calculated the same measures as show in table 2

**Table 2:** The structural properties of the communities formed by social network, being <k> medium grade, (T) transitivity, (A) assortativity, <B> Average betweenness bentrality, (D) medium way distance, (N) vertex and (E) edges.

| Communities | <k> | T | A | <B> | D | N | E | $S_1$ |
|---|---|---|---|---|---|---|---|---|
| 1 | 39 | 0,894 | -0,5567 | 67,47 | 1,896 | 331 | 6595 | 74,43 |
| 2 | 64 | 0,958 | -0,5355 | 61,82 | 0,997 | 77 | 2494 | 24,94 |
| 3 | 67 | 0,829 | -0,496 | 502,13 | 1,574 | 151 | 5105 | 83,85 |
| 4 | 21 | 0,989 | -0,4294 | 22,27 | 1,201 | 28 | 305 | 25,49 |
| 5 | 28 | 0,966 | -0,3671 | 362,93 | 1,062 | 31 | 436 | 9,68 |

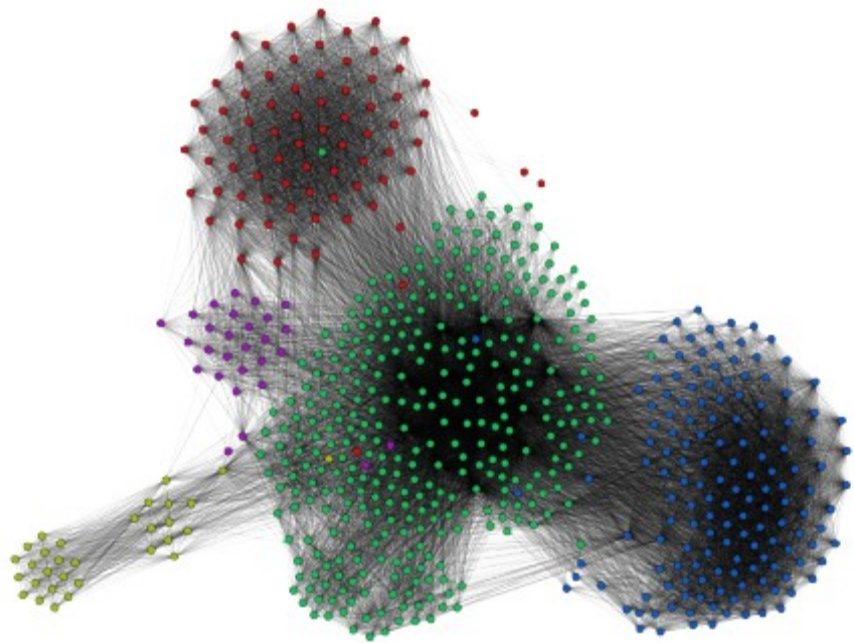

**Figure 3:** social network with communities calculated from de books The Silmarillion, The Hobbit and The Lord of The Rings (unic volume).

Table 3 shows the measurements of the structural properties and the entropy by community. After the withdrawal of deities we can observe the fall of entropy values, as we also see in the comparison of Figure 4.

**Table 3:** The structural properties of the communities formed by the social network with the removal of the deities, being <k> the medium degree, (T) the transitivity, (A) the assertiveness, <B> the medium betweenness centrality (d) average path length, (N) vertices and (E) edges.

| Communities | <k> | T | A | <B> | D | N | E | $S_1$ |
|---|---|---|---|---|---|---|---|---|
| 1 | 65 | 0,657 | -0,465 | 42,161 | 1,569 | 149 | 5051 | 32,602 |
| 2 | 66 | 0,997 | -0,068 | 0,906 | 1,062 | 32 | 436 | 8,134 |
| 3 | 26 | 0,996 | -0,027 | 4,828 | 1,128 | 76 | 2493 | 14,621 |
| 4 | 19 | 0,541 | -0,071 | 38,918 | 1,819 | 98 | 1306 | 23,249 |
| 5 | 28 | 0,167 | -0,566 | 122,059 | 1,972 | 252 | 2419 | 8,435 |

The figure 4 illustrates the behavior of the calculated entropy with the formalism of the equation 10.

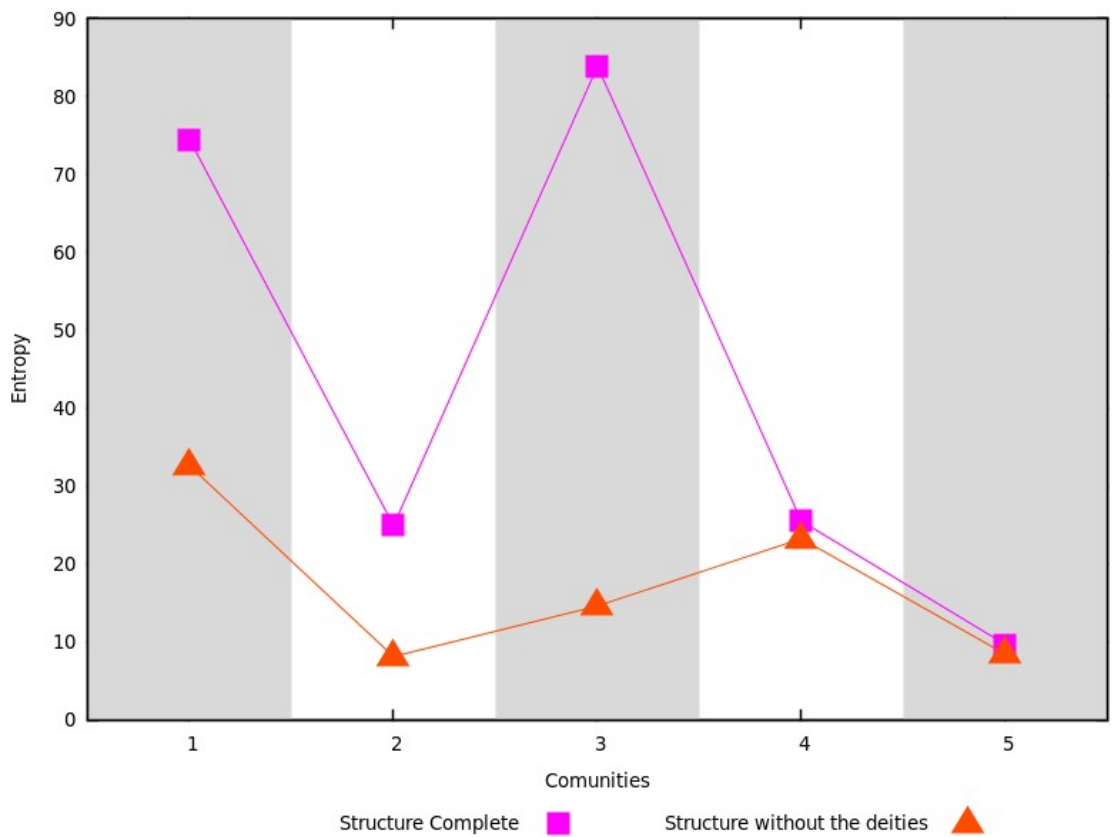

**Figure 4:** Comparative graphic of the structural entropies considering the bond of de sequence of the formed communities (subgraphs). We can observe the behavior of the entropy when the deities are extracted from the network.

Communities showed high transitivity values, indicating high density in links. This density is characterized by the number of triangles formed by the social relation of sharing event and allusion to the social relation. However, communities presented themselves as negative about the assortativity, which denotes vertices with high connection linking to the vertices with low connection and the presence of hubs in the structure. As can be observed in the community of the creator gods and all members linked directly to them. This community presented higher average degree and higher average betweenness centrality. This happens because the reference of allusion with the deities and the other inhabitants of Middle Earth, increase the connection among the characters. An example is the elf Elrond that is present in all works. Being one of the intermediate vertices between the action of the divine characters and not divine in the Tolkien allegory. The characters with this characteristic resulted in heterogeneity among the links. Among these communities, there is the community 3 introducing as the most heterogeneous in connections because the structural entropy calculated takes as link the sequence of the social structure degree. Thus, the structure of the works of J.R.R. Tolkien is dense in connections and their communities are heterogeneous with the bond of degree sequence.The average path length of the network communities showed that the paths taken in the structure between the characters it is at most 1.8, in other words, a character takes distance at least approximately from two edges of another character. This occurs by the relation of allusion and sharing of events. In the plot of the works this is reinforced by the wars that happens and the reports of the ancestors, because each character of Tolkien has a history and a genealogy

# 3 Methodology

To represent and analyze mathematically we will use the theory of the graphs. A graph is defined as an ordinate pair that relates with a group of vertices (V) and a group of edges (E) [16], [17]. The edges may or may not have a direction, what makes a graph directed or not, respectively. This form of presentation of a social network helps in the calculation of the main measures of centrality to characterize the formed structures.

### 3.1 Degree centrality and its distribution

The degree centrality measure constitutes from the number of bonds that the vertice i realizes with the next vertice j, so, it is defined as a vertice group cardinality of the group of vertices next to i. In that measure, we can extract two quantities. The first is the medium degree that constitutes the arithmetic medium of de degree sequence.

$$\langle k \rangle = \frac{\sum_{i=1}^{N} k_i}{N} \qquad (1)$$

If N is the number of vertices and the degree from the vertice i. Other possible measure is the distribution of degree, being the fraction of vertices with degree k. This distribution enables in a simple way to qualify the behavior of the bonds in the structures [18]. The model proposed to the adjustment of distribution of degree in this work is given by:

$$p(k|\alpha, \beta, \gamma, \nu) = \nu \frac{e^{-k\beta}}{(k+\alpha)^{\gamma}} \qquad (2)$$

Being γ the decay parameter, α the displacement parameter, β the exponential factor cut parameter and ν the normalization and distribution parameter. This distribution corrects the anomaly to vertices with high probability of connections.

### 3.2 Betweeness centrality

The betweeness centrality is a measure of influence of the vertice i. This centrality measure is based on the minor paths, or geodetics that pass by i[19]. If σ(i, j) is the number of geodetic paths between the vertices i and j and the number of these that cross vertice l that intermediates the vertices i and j, then the Betweenness centrality can be defined as:

$$g_l = \frac{2}{(N-1)(N-2)} \sum_{i \neq j} \frac{\sigma_l(i,j)}{\sigma(i,j)} \qquad (3)$$

### 3.3 Clustering coefficient

The clustering coefficient or transitivity is a measure that quantifies the number of triangles on the [20],[21]. Being calculated as:

$$C_i = \frac{2n_i}{k_i(k_i - 1)} \qquad (4)$$

We can also define as the probability measure of a vertice i to be connected with an vertice j. Then we define the medium transitivity as:

$$\langle C \rangle = \frac{\sum C_i}{N} \qquad (5)$$

### 3.4 Assortativity

This measure if defined by the connection preference between the vertices. This coefficient can assume values between [-1,1]. In which, positive values of (A) indicate that the vertices of similar degrees are connected, characterizing a network as assortative. For negative values, vertices with the elevated degree connect with vertices with low degree, characterizing the network as disassortative [22]. This value is given by:

$$A = \frac{\sum_i e_{ii} - \sum_i a_i b_i}{1 - \sum_i a_i b_i} \qquad (6)$$

Being $e_{ii}$ the fraction of the edges that connect the vertices i and j, $a_i$, $\sum_i e_{ij}$ e $b_j$ $\sum_j e_{ij}$.

### 3.5 Communities

The community structure is defined as participations from a G graph. These participations are obtained with the modularity calculi, give by:

$$Q = \sum_i \left[ e_{ii} - \left( \sum_j e_{ii} \right)^2 \right] \qquad (7)$$

Being $e_{ii}$ and $e_{ij}$ the edge fractions in the network that connect the group i from this group j. The edge fractions are the one which fit among the communities [23].

### 3.6 Average path length

The average path length of a graph is the measurement of the distances between all pairs of vertices. So, if d (*u,v*) is the distance between vertices *u* and *v*, N the number of vertices in the graph, so:

$$d = \frac{\sum_{u,v \in V} d(u, v)}{\binom{N}{2}} \qquad (8)$$

### 3.7 Structural Entropy

The complexity of a structure of network depends on the global organization that the bonds make with their elements. The real networks present different levels of organization, that is because of the heterogeneity of the bonds among pairs of vertices. Therefore, to measure the organization degree or complexity of a network we utilize the definition of entropy from the statistic mechanic.

This entropy is calculated given the structural bonds of the networks, for example the degree sequence [24] [25]. This structural bond defines the number of bonds that the vertice I realizes with its adjacent. In this case, we have the participation function for this system, given by:

$$Z_1 = \sum_{a_{ij}} \prod_i \delta \left( k_i - \sum_{a_{ij}} \right) exp \left[ \sum_{i<j} h_{ij} a_{ij} \right] \qquad (9)$$

The approximation for the entropy from this ensemble is given by the Lagrange's multipliers method to optimize the equation 9, therefore, the entropy of order one ($s_1 = S_1/N$) by the vertice of the network is given by:

$$s_1 \approx -\frac{1}{N} \left[ \sum k_i (\ln(k_i) - 1) - \frac{1}{2} \sum_i \ln(2\pi k_i) + \frac{1}{2} \langle k \rangle N [\ln(\langle k \rangle N) - 1] - \frac{1}{2} \left( \frac{\langle k^2 \rangle}{\langle k \rangle} \right)^2 \right] \qquad (10)$$

Where N the number of the vertices of the network, $k_i$ is the degree of the vertice I, <k> is the medium degree.

## 4 Formation of the networks

To form the literary social network were utilized the works (The Silmarillon, The Hobbit and The Lord of The Rings Single Volume) from the writer J.R.R. Tolkien. The mining of the characters from the works was considering the social restrictions of allusion, shared events and direct interaction, which formed the bonds between the characters. This division to the mining is given in three parts.

1. Select in the text the characters that are participating of the plot of the book;

2. Analysis of the kind of connection given by dialogues, events and description. These connections must respect the allusion restrictions, shared events and direct interaction;

3. Formation of a pattern of the social network kind given the steps 1 and 2.

The figure 5 shows the mining of the characters in one part of the book "The Fellowship of the Ring" which is part of Elrond's council.

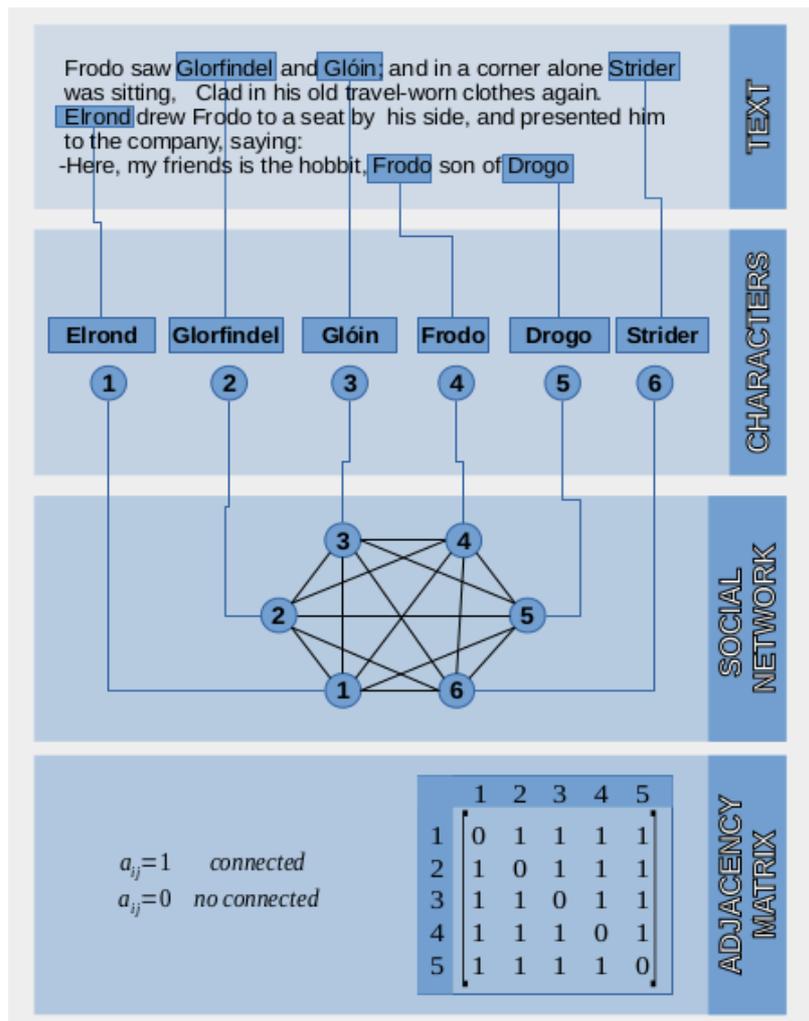

**Figure 5:** Scheme of the character mining from a part of the book The Fellowship of the Ring. TEXT: identifies the characters in the text. CHARACTERS: the names of the characters are separated and at SOCIAL NETWORK the three social relations are analyzed and will be bonds between the characters, forming then a literary social network. ADJACENCY MATRIX represent the matrix form of a graph.

## 5 Conclusion

The purpose of this article was to construe a social network of The Lord Of The Rings, The Silmarillion and The Hobbit. The social network presented itself as disassortative, in other words, vertices with high connection relate with vertices that has lower connectivity and with transitivity of approximately 0,30. This transitivity reflects in the formation of triangles on the network, characterizing its density. The distribution of the network degree behaved as a distribution of the type potentiates law with truncation exponential, as given in equation 2. The setting was satisfact as shown the coefficient of Pearson. In this network were found 5 communities with the Walktrap Communities algorithm, that optimizes tue function of modularity by aleatory walk. The statistics of the communities showed that the community 2 has a centrality of higher average betweenness centrality. This happens due to the relations of allusion to to the relations of friendship and sharing events that the divine beings has on the storyline of the works.

High density connections as shown in the transitivity measure, is due to sharing of events and direct edges. All communities have an disassortative character as verified in the whole structure. With the calculation of structural entropy, it can be quantify the complexity of the structures formed by communities, in which stands out more the community 3, where is presented the divine beings and mythological creatures next to them.

The emerging structure of Tolkien's works presented one disassortative character, transitive and with a distribution of degree with social networking characteristics.

The interaction of the gods to the characters happened in a way that the average path length having D= 1.988, with the extraction of deities the measure passes to D= 3.961, that second [7] shows the omnipresent character of deities and thereby lowering the structural entropy.